\documentclass[aps,10pt,pre,twocolumn]{revtex4-1}

\usepackage{amsmath,amssymb,amsfonts,amsthm}
\usepackage[utf8]{inputenc}
\usepackage{graphicx}
\usepackage[caption=false]{subfig}
\usepackage{bbm}
\usepackage[pdftex,bookmarks=false,colorlinks=true,linkcolor=blue,
citecolor=blue,filecolor=black,urlcolor=blue]{hyperref}

\providecommand{\ud}{d}
\providecommand{\abs}[1]{\left\lvert#1\right\rvert}

\begin{document}

\title{Matrix-valued Boltzmann equation for the non-integrable Hubbard chain}

\author{Martin L.R. F\"urst}
\email{mfuerst@ma.tum.de}
\affiliation{Excellence Cluster Universe,
Technische Universit\"at M\"unchen,
Boltzmannstra{\ss}e 2}
\affiliation{Mathematics Department,
Technische Universit\"at M\"unchen,
Boltzmannstra{\ss}e 3,
85748 Garching bei M\"unchen,
Germany}

\author{Christian B. Mendl}
\email{mendl@ma.tum.de}
\affiliation{Mathematics Department,
Technische Universit\"at M\"unchen,
Boltzmannstra{\ss}e 3}

\author{Herbert Spohn}
\email{spohn@ma.tum.de}
\affiliation{Mathematics Department,
Technische Universit\"at M\"unchen,
Boltzmannstra{\ss}e 3}
\affiliation{Physics Department,
Technische Universit\"at M\"unchen,
James-Franck-Stra{\ss}e 1}

\date{\today}

\begin{abstract}
The standard Fermi-Hubbard chain becomes non-integrable by adding to the nearest neighbor hopping additional longer range hopping amplitudes. We assume that the quartic interaction is weak and investigate numerically the dynamics of the chain on the level of the Boltzmann type kinetic equation. Only the spatially homogeneous case is considered. We observe that the huge degeneracy of stationary states in case of nearest neighbor hopping is lost and the convergence to the thermal Fermi-Dirac distribution is restored. The convergence to equilibrium is exponentially fast. However for small n.n.n.\ hopping amplitudes one has a rapid relaxation towards the manifold of quasi-stationary states and slow relaxation to the final equilibrium state.
\end{abstract}

\maketitle


\section{Introduction}
\label{sec:Introduction}

The most widely known quantum chains are \emph{integrable}, in the sense that they have a large number of local conservation laws. Eigenfunctions can be determined through the Bethe ansatz and there are special relations for scattering amplitudes, to mention only a few characteristics, see~\cite{Essler2010,HubbardModelPhysics1995} for further details. Obviously, dynamical properties depend sensitively on the integrable structure. For example,  such chains  have a large Drude weight generically, signaling ballistic transport but still leaving room for a diffusive component~\cite{DiffusionBallisticTransport2009}. There has been considerable efforts to understand what happens as  one moves away from integrability~\cite{ConservationIntegrabilityTransport2011,QuantumLiquids2012}. In our contribution we study the case where integrability is lost by adding couplings beyond the nearest neighbor ones. But we will stay in the regime where kinetic theory remains applicable. More than by other methods, we arrive at detailed information on how non-integrability becomes manifest.

Specifically we consider the Fermi-Hubbard chain with hamiltonian
\begin{equation}
\label{eq:Hamiltonian}
H = \sum_{x, y \in \mathbb{Z}} \alpha(x - y) \, a(x)^* \cdot a(y) + \frac{\lambda}{2} \sum_{x \in \mathbb{Z}} \big( a(x)^* \cdot a(x) \big)^2
\end{equation}
%
%
with $a(x)^* \cdot a(x) = a_\uparrow(x)^* \, a_\uparrow(x) + a_\downarrow(x)^* \, a_\downarrow(x)$.  $\alpha(x)$ is the hopping amplitude, satisfying $\alpha(x) = \alpha(x)^*$, $\alpha(x) = \alpha(-x)$, and $\lambda$ is the strength of the on-site interaction.

$H$ is integrable for the nearest neighbor hopping amplitude, i.e.  $\alpha(\pm 1) = 1$, $\alpha(x) = 0$ otherwise. For longer range hoppings $H$ is commonly expected to be non-integrable. On the kinetic level, changing $\alpha$ amounts to changing the dispersion relation. Otherwise the structure of the transport equation is not altered. Thus the issue of non-integrability is fairly accessible to the Boltzmann kinetic equation.

One aspect was studied in detail already in~\cite{BoltzmannFermi2012}, where it was noted that for nearest neighbor coupling the Hubbard-Boltzmann equation has a much larger set of stationary solutions than usually anticipated. On the other hand for the domain of attraction of a non-thermal stationary state, the usual kinetic picture is valid. Entropy is strictly increasing and the steady state is approached exponentially fast. (We always work in the spatially homogeneous set-up.) Our goal here is to study the approach to stationarity once $\alpha$ is no longer of nearest neighbor type. As in~\cite{BoltzmannFermi2012} we will rely on numerical solutions of the Boltzmann-Hubbard equation and study two prototypical non nearest neighbor hoppings.
\medskip\\
(i) An additional next-nearest neighbor hopping term, i.e., $\alpha(0) = 1$, $\alpha(\pm 1) = -\tfrac{1}{2}$, $\alpha(\pm 2) = -\tfrac{\eta}{2}$, and $\alpha(x) = 0$ otherwise, with a tunable parameter $\eta \in \mathbb{R}$. The Fourier transform of $\alpha$ is the dispersion relation
\begin{equation}
\label{eq:omega_nnn}
\omega_{\eta}(k) = 1 - \cos(2 \pi k) - \eta \cos(4 \pi k).
\end{equation}
The nearest neighbor case corresponds to $\eta = 0$.
\medskip\\
(ii) An exponential decay of higher-order hopping terms, i.e., $\alpha(0) = -1$, $\alpha(x) = -\tfrac{1}{2} \mathrm{e}^{-\zeta \abs{x}}$ for $x \neq 0$, with a tunable parameter $\zeta > 0$. 
The Fourier transform of $\alpha$ is the dispersion relation
\begin{equation}
\label{eq:omega_exp}
\omega_{\zeta}(k) = -\sum_{j=0}^{\infty} \mathrm{e}^{-\zeta j} \cos(2\pi j k).
\end{equation}
The limit $\zeta \to \infty$ corresponds to the nearest neighbor case after shifting and rescaling $\mathrm{e}^{\zeta}(1 + \omega_{\zeta}(k))$, while $\zeta \to 0$ allows for large hoppings of size $1/\zeta$.

\begin{figure}[!ht]
\centering
\includegraphics[width=0.8\columnwidth]{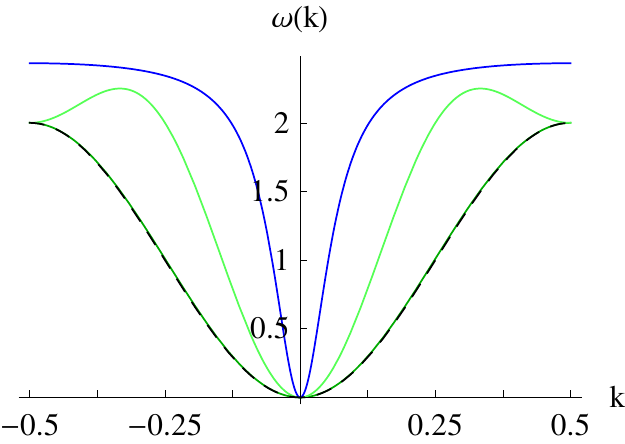}
\caption{(Color online) The dispersion relation $\omega(k)$ for the next-nearest neighbor model in Eq.~\eqref{eq:omega_nnn} with $\eta_1 = \frac{1}{200}$ and $\eta_2 = \frac12$ (green solid curves coinciding with the dashed line and with 2 local maxima, respectively), and for the exponential hopping model in Eq.~\eqref{eq:omega_exp} with $\zeta = \frac{2}{5}$ (upper blue solid curve). All curves are shifted such that $\omega(0) = 0$. The dashed curve shows the dispersion relation for the (reference) nearest neighbor hopping model.}
\label{fig:Omega}
\end{figure}

Fig.~\ref{fig:Omega} visualizes $\omega(k)$ for both cases i) and ii), as well as the ``reference'' nearest neighbor hopping model (black dashed line). Later the next-nearest neighbor hopping model is investigated numerically for a small $\eta_1 = \frac{1}{200}$ (dark green line in Fig.~\ref{fig:Omega}) as well as $\eta = \frac12$ (light green line).

Our goal is to study the dynamics of the Hubbard chain at small interaction in dependence on $\eta$, respectively $\zeta$. For this purpose, in Section~\ref{sec:Bequation} we first recall the structure of the corresponding Boltzmann transport equation. The collision rules for quasiparticles are implicitly determined by conservation of momentum and energy, which will be discussed in Section~\ref{sec:Collisions}. The numerical scheme is explained in Section~\ref{sec:Numerics}, which is the technical backbone of our investigations. We this tool we study the approach to a stationary state, see Section~\ref{sec:Results}, and its dependence on the collision rules, in other words on the dispersion relation.

In~\cite{TransportWeaklyInteracting2012} mass diffusion in dependence on $\eta$ was studied for a ``toy'' linear transport equation. The divergence of this transport coefficient as $\eta \to 0$ is related to our findings for the full Boltzmann equation.

\section{The Boltzmann-Hubbard Equation}
\label{sec:Bequation}

We briefly recall the structure of the Boltzmann-Hubbard equation, see~\cite{BoltzmannFermi2012} for details. For the Fourier transformation we use the convention
\begin{equation}
\label{eq:Fouriertransformation}
\hat{f}(k) = \sum_{x \in \mathbb{Z}} f(x)\, \mathrm{e}^{-2 \pi i\, k \cdot x}.
\end{equation}
Then the first Brillouin zone is the interval $\mathbb{T} = [-\tfrac{1}{2}, \tfrac{1}{2}]$ with periodic boundary conditions. The dispersion relation $\omega(k) = \hat{\alpha}(k)$ and, up to a constant, in Fourier space $H$ can be written as
\begin{equation}
\begin{split}
H &= \sum_{\sigma \in \{\uparrow,\downarrow\}} \int_{\mathbb{T}} \ud k \, \omega(k)\, \hat{a}_\sigma(k)^*\, \hat{a}_\sigma(k) \\
&\quad + \frac{\lambda}{2} \int_{\mathbb{T}^4} \ud^4 \boldsymbol{k} \, \delta(\underline{k}) \, \hat{a}_\uparrow(k_1)^*\, \hat{a}_\uparrow(k_2)^*\, \hat{a}_\downarrow(k_3)\, \hat{a}_\downarrow(k_4)
\end{split}
\end{equation}
with $\underline{k} = k_1 + k_2 - k_3 - k_4 \mod 1$ and $\ud^4 \boldsymbol{k} = \ud k_1\, \ud k_2\, \ud k_3\, \ud k_4$.

To arrive at the kinetic equation, we assume that the initial state of the chain is quasifree, gauge invariant, and invariant under spatial translations. It is thus completely characterized by the two-point function
\begin{equation} \label{eq:InitialCondition}
\langle \hat{a}_\sigma(k)^*\, \hat{a}_\tau(k') \rangle = \delta(k - k') W_{\sigma \tau}(k).
\end{equation}
It will be convenient to think of $W(k)$ as a $2 \times 2$ matrix for each $k \in \mathbb{T}$. Then, in general, $W(k_1) W(k_2) \neq W(k_2) W(k_1)$ and every argument of standard kinetic theory has to be reworked. By the Fermi property we have $0 \leq W(k) \leq 1$ as a matrix for each $k$. In particular, $W$ can be written as
\begin{equation}
W(k) = \sum_{\sigma \in \{\uparrow,\downarrow\}} \varepsilon_\sigma(k) \lvert k,\sigma\rangle \langle k,\sigma\rvert,
\end{equation}
where $\lvert k,\sigma\rangle$ for $\sigma \in \{\uparrow,\downarrow\}$ is a $k$-dependent basis in spin space $\mathbb{C}^2$ and $\varepsilon_\sigma$ are the eigenvalues with $0 \leq \varepsilon_\sigma \leq 1$.

At some later time $t$ the state is still gauge and translation invariant, hence necessarily
\begin{equation}
\langle a_\sigma(k,t)^*\, a_\tau(k',t) \rangle = \delta(k - k') W_{\sigma \tau}(k,t).
\end{equation}
In general $W(t)$ is a complicated object, but for small coupling $\lambda$ the quasi-free property persists over a time scale of order $\lambda^{-2}$, a structure which allows one to obtain the kinetic equation by second order time-dependent perturbation theory. More details can be found, e.g., in~\cite{ErdosSalmhoferYau2004,NotToNormalOrder2009,MeiLukkarinenSpohnPrep}. Here we only write down the resulting Boltzmann equation
\begin{equation}
\label{eq:BoltzmannEquation}
\frac{\partial}{\partial t} W(k,t)
= \mathcal{C}_\mathrm{c}[W](k,t) + \mathcal{C}_\mathrm{d}[W](k,t) = \mathcal{C}[W](k,t),
\end{equation}
which has the structure of an evolution equation and has to be supplemented with the initial data $W(k,0) = W(k)$.

The first term is of Vlasov type,
\begin{equation}
\label{eq:Cc}
\mathcal{C}_\mathrm{c}[W](k,t) = - i\,[H_\mathrm{eff}(k,t), W(k,t)],
\end{equation}
where the effective hamiltonian $H_\mathrm{eff}(k,t)$ is a $2 \times 2$ matrix which itself depends on $W$. More explicitly,
\begin{multline}
\label{eq:Heff}
H_{\mathrm{eff},1} = \int_{\mathbb{T}^3} \ud k_2 \ud k_3 \ud k_4 \, \delta(\underline{k}) \, \mathcal{P} \left(\tfrac{1}{\underline{\omega}}\right)\\
\times \big( W_3 W_4 - W_2 W_3 - W_3 W_2 - \mathrm{tr}[W_4] W_3 + \mathrm{tr}[W_2] W_3 + W_2 \big).
\end{multline}
Here and later on we use the shorthand $\tilde{W} = 1 - W$, $W_1 = W(k_1,t)$, $H_{\mathrm{eff},1} = H_\mathrm{eff}(k_1,t)$, $\underline{\omega} = \omega(k_1) + \omega(k_2) - \omega(k_3) - \omega(k_4)$. Since $W$ is $2 \times 2$ matrix-valued, $\mathrm{tr}[\,\cdot\,]$ is the trace in spin space. Finally $\mathcal{P}$ denotes the principal part. Since the $k_3$, $k_4$ integration can be interchanged, $H_{\mathrm{eff}} = H_{\mathrm{eff}}^*$, as it should be.

There are many different ways to write the collision term $\mathcal{C}_\mathrm{d}$. We choose a version which separates the various contributions into gain and loss term. Then
\begin{multline}
\label{eq:Cd}
\mathcal{C}_\mathrm{d}[W]_1 = \pi \int_{\mathbb{T}^3} \ud k_2 \ud k_3 \ud k_4 \, \delta(\underline{k}) \, \delta(\underline{\omega}) \\
\times \big( \mathcal{A}[W]_{1234} + \mathcal{A}[W]_{1234}^* \big),
\end{multline}
where the index $1234$ means that the matrix $\mathcal{A}[W]$ depends on $k_1$, $k_2$, $k_3$, and $k_4$. Explicitly
\begin{multline}
\mathcal{A}[W]_{1234} = - W_4 \tilde{W}_2 W_3 + W_4 \, \mathrm{tr}[ \tilde{W}_2 W_3 ] \\
- \big\{\tilde{W}_4 W_3 - \tilde{W}_4 W_2 - \tilde{W}_2 W_3 + \tilde{W}_4 \, \mathrm{tr}[ W_2 ] \\
- \tilde{W}_4 \, \mathrm{tr}[ W_3 ] + \mathrm{tr}[ W_3 \tilde{W}_2 ] \big\} W_1
\end{multline}
with the first two summands the gain term and $\{ ... \} W_1$ the loss term. The gain term is always positive definite, as implied by the inequality
\begin{equation}
A \, \mathrm{tr}[B C] + C \, \mathrm{tr}[B A] - A B C - C B A \geq 0
\end{equation}
valid for arbitrary positive definite matrices $A, B, C$. Thus if an eigenvalue of $W(k,t)$ happens to vanish, the gain term pushes it back to values $> 0$. A similar argument can be made for $\tilde{W}(k,t)$, implying the propagation of the Fermi property~\cite{MeiLukkarinenSpohnPrep}, to say: if at $t = 0$ one has $0 \leq W(k) \leq 1$, then the solution to~\eqref{eq:BoltzmannEquation} also satisfies $0 \leq W(k,t) \leq 1$.

In general, ''spin'',
\begin{equation}
\label{eq:SpinConservation}
\int_{\mathbb{T}^d} \ud k \, W(k,t)
\end{equation}
and energy
\begin{equation}
\label{eq:EnergyConservation}
\int_{\mathbb{T}^d} \ud k \, \omega(k) \, \mathrm{tr}[W(k,t)]
\end{equation}
are conserved. In the long time, $W(k,t)$ will become diagonal in the conserved spin basis. Each component has a Fermi-Dirac distribution with common temperature and destined chemical potentials, which then is precisely in accordance with the parameters from the conservation laws.


For the nearest neighbor model, one has the additional conservation law 
\begin{equation}
\label{eq:TraceConservation}
\frac{\ud}{\ud t} \left( \mathrm{tr}[W(k,t)] - \mathrm{tr}[W(\tfrac12-k,t)] \right) = 0.
\end{equation}
All stationary states are necessarily of the form
\begin{equation}
W_{\mathrm{nth}}(k) = \sum_{\sigma \in \{\uparrow,\downarrow\}} \left( \mathrm{e}^{f(k) - a_\sigma} + 1 \right)^{-1} \lvert\sigma\rangle\langle\sigma\rvert,
\end{equation}
with $f(k) = -f(\frac{1}{2} - k)$. $W_{\mathrm{nth}}$ is an equilibrium state if $f(k) = \beta \omega(k)$.

The entropy of the state $W$ is then defined by
\begin{equation}
S[W] = - \int_{\mathbb{T}^d} \ud k_1 \big( \mathrm{tr}[W_1 \log W_1] + \mathrm{tr}[\tilde{W}_1 \log \tilde{W}_1] \big).
\end{equation}
in accordance with an ideal Fermi gas. It is easily checked that the entropy production $\sigma \geq 0$,
\begin{equation}
\label{eq:sigmaW}
\begin{split}
\sigma[W] &= \frac{\ud}{\ud t} S[W] \\
&= -\int_{\mathbb{T}^d} \ud k_1 \, \mathrm{tr}[ (\log W_1 - \log \tilde{W}_1) \, \mathcal{C}[W]_1 ].
\end{split}
\end{equation}
The H-theorem asserts that
\begin{equation}
\label{eq:HTheorem}
\sigma[W] \geq 0 \qquad \text{for all } W \text{ with } 0 \leq W \leq 1.   
\end{equation}


\section{Collisions}
\label{sec:Collisions}

\subsection{Next-nearest neighbor model}

\begin{figure}[b]
\includegraphics[width=0.9\columnwidth]{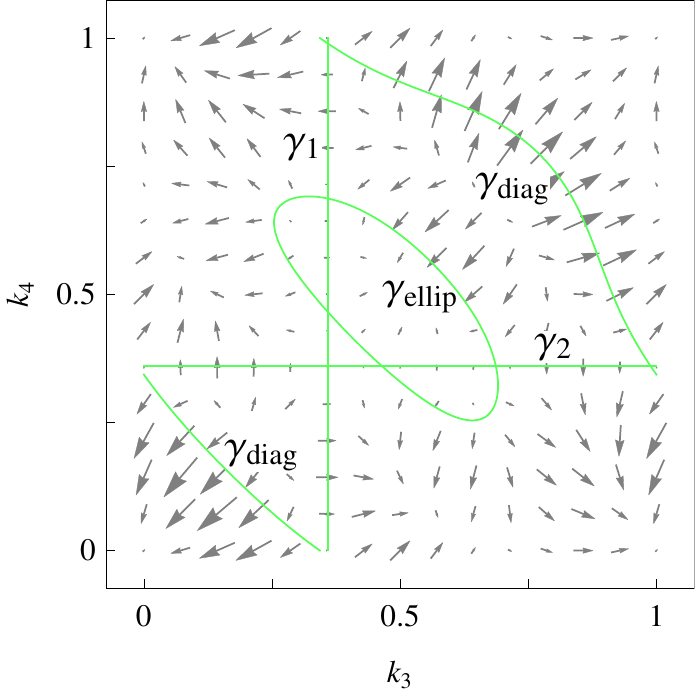}
\caption{(Color online) Contour (green straight lines) and gradient (gray vectors) of the next-nearest neighbor energy conservation contour $\underline{\omega}_{\eta} = 0$ (with $\eta = \frac12$) for fixed $k_1 = \tfrac{23}{64}$ and after eliminating $k_2$. The vertical and horizontal lines, $\gamma_1$ and $\gamma_2$, are the contours $k_3 = k_1$ and $k_4 = k_1$, respectively. The contour $\gamma_{\mathrm{ellip}}$ disappears when $\abs{\eta} < \frac{1}{4}$.}
\label{fig:OmegaNNNEcons}
\end{figure}

The starting point is to investigate the kinematically allowed collisions $\delta(\underline{k})\delta(\underline{\omega}_{\eta})$. Using momentum conservation $\underline{k} = 0 \mod 1$ and defining $s_{12} = k_1 + k_2 \equiv k_3 + k_4$, $\Delta k_{12} = \frac{1}{2}(k_1 - k_2)$ and $\Delta k_{34} = \frac{1}{2}(k_3 - k_4)$, one arrives at the factorization
\begin{equation}
\label{eq:EnergyConservationFactorized}
\underline{\omega}_{\eta} = \underline{\omega}_{\mathrm{bas}} \ \underline{\omega}_{\mathrm{add},\eta}
\end{equation}
with the factors
\begin{equation}
\label{eq:omega_bas}
\begin{split}
\underline{\omega}_{\mathrm{bas}}
&= 4 \sin(\pi(k_1 - k_3)) \sin(\pi(k_1 - k_4))\\
&= 2 \left(\cos(2\pi \Delta k_{34}) - \cos(2\pi \Delta k_{12})\right)
\end{split}
\end{equation}
and
\begin{equation}
\label{eq:omega_add_eta}
\begin{split}
\underline{\omega}_{\mathrm{add},\eta}
&= \cos(\pi\,s_{12}) + \eta\,\cos(2\pi\,s_{12})\\
& \quad \times \left(\cos(2\pi\,\Delta k_{12}) + \cos(2\pi\,\Delta k_{34})\right).
\end{split}
\end{equation}
Eq.~\eqref{eq:EnergyConservationFactorized} is of similar form as~\cite[Eq.~(34)]{BoltzmannFermi2012}, except for the additional $\eta$-dependent term in $\underline{\omega}_{\mathrm{add},\eta}$. In particular, the ``trivial'' solution paths $k_3 = k_1$ (denoted $\gamma_1$) and $k_4 = k_1$ (denoted $\gamma_2$) remain unaffected by $\eta$. A sign change of $\eta$, i.e., $\eta \to -\eta$, corresponds to $k_i \to k_i + \frac12$ since this transformation sends $s_{12} \to s_{12} + 1$ and $\cos(\pi\,s_{12}) \to -\cos(\pi\,s_{12})$, while the other cosine terms in Eq.~\eqref{eq:omega_add_eta} are unaffected. Thus without loss of generality one may assume that $\eta \ge 0$.

We decompose
\begin{equation}
\mathcal{A}[W]_{1234} + \mathcal{A}[W]_{1234}^* = \mathcal{A}_{\mathrm{quad}}[W]_{1234} + \mathcal{A}_{\mathrm{tr}}[W]_{1234}
\end{equation}
with
\begin{multline}
\mathcal{A}_{\mathrm{quad}}[W]_{1234} = -\tilde{W}_1 W_3 \tilde{W}_2 W_4 - W_4 \tilde{W}_2 W_3 \tilde{W}_1\\
+ W_1 \tilde{W}_3 W_2 \tilde{W}_4 + \tilde{W}_4 W_2 \tilde{W}_3 W_1, \label{eq:Aquad}
\end{multline}
\begin{multline}
\mathcal{A}_{\mathrm{tr}}[W]_{1234} = \big(\tilde{W}_1 W_3 + W_3 \tilde{W}_1\big) \mathrm{tr}[\tilde{W}_2 W_4]\\
- \big(W_1 \tilde{W}_3 + \tilde{W}_3 W_1\big) \mathrm{tr}[W_2 \tilde{W}_4].
\label{eq:Atr}
\end{multline}
The discussion of the integration along $\gamma_1$, $\gamma_2$ follows the same line as in~\cite{BoltzmannFermi2012}: $\mathcal{A}_{\mathrm{quad}}$ is zero along both $\gamma_1$, $\gamma_2$, but $\mathcal{A}_{\mathrm{tr}}$ is zero along $\gamma_1$ only.

Concerning the factor $\underline{\omega}_{\mathrm{add},\eta}$ in Eq.~\eqref{eq:EnergyConservationFactorized}, the contour $\underline{\omega}_{\mathrm{add},\eta} = 0$ splits into two parts, denoted $\gamma_{\mathrm{diag}}$ and $\gamma_{\mathrm{ellip}}$, see Fig.~\ref{fig:OmegaNNNEcons}. Using the identity $\cos(2\pi\,s_{12}) = 2 \cos^2(\pi\,s_{12}) - 1$ and solving for $s_{12}$, one arrives at
\begin{equation}
\label{eq:s12}
s_{12}(r) = \frac{1}{\pi}\, \mathrm{arccos}\!\left(\frac{\pm \sqrt{1+2\,r^2}-1}{2\,r}\right)
\end{equation}
for $\gamma_{\mathrm{diag}}$ and $\gamma_{\mathrm{ellip}}$, respectively, where
\begin{equation}
\label{eq:r}
r = 4\,\eta\left(\cos(2\pi\,\Delta k_{12}) + \cos(2\pi\,\Delta k_{34})\right).
\end{equation}
The argument of the $\mathrm{arccos}$ function in Eq.~\eqref{eq:s12} should be in the interval $[-1,1]$, which is always satisfied for $\gamma_{\mathrm{diag}}$. However, on $\gamma_{\mathrm{ellip}}$ this constraint leads to the condition $\abs{r} \ge 2$. Thus we conclude that the contour $\gamma_{\mathrm{ellip}}$ disappears for $\abs{\eta} < \frac{1}{4}$ since by Eq.~\eqref{eq:r}, $\abs{r} \le 4 \abs{\eta} 2 < 2$.

As a remark, Taylor-expansion at $r = 0$ of Eq.~\eqref{eq:s12} on $\gamma_{\mathrm{diag}}$ gives
\begin{equation}
s_{12}(r) = \frac{1}{2} - \frac{r}{2 \pi} + \frac{11 r^3}{48 \pi} - \dots
\end{equation}
In particular, we reobtain the constant $s_{12} \equiv \frac12$ for the next-neighbor case $\eta = 0$.

In summary, the contour $\gamma_{\mathrm{diag}}$ is deformed as compared to the nearest neighbor case (compare with~\cite[Fig.~2]{BoltzmannFermi2012}). The additional collision channel $\gamma_{\mathrm{ellip}}$ appears when $\abs{\eta} \ge \frac{1}{4}$. The gradient vector field of $\underline{\omega}_{\eta}$ (gray vectors in Fig.~\ref{fig:OmegaNNNEcons}) is noticeable different compared to the nearest neighbor case.

\subsection{Exponential hopping}

\begin{figure}[b]
\includegraphics[width=0.9\columnwidth]{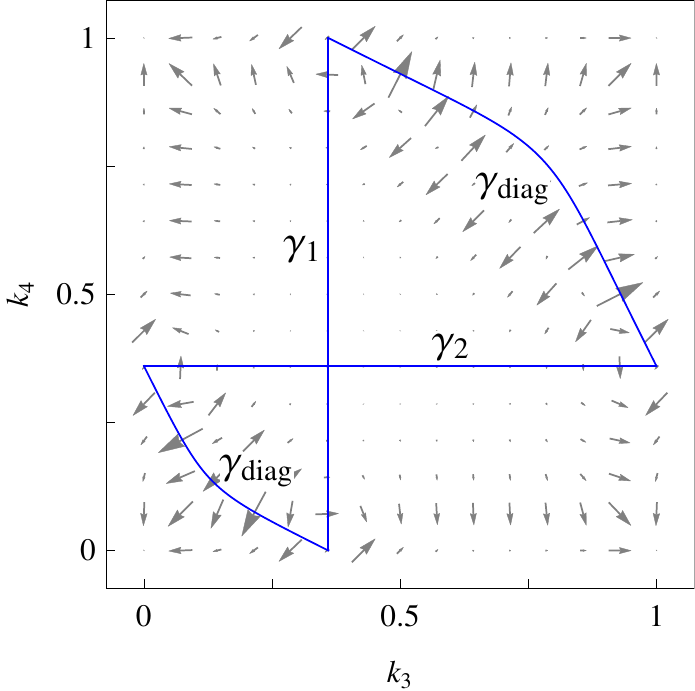}
\caption{(Color online) Contour (blue straight lines) and gradient (gray vectors) of the exponential decay energy conservation $\underline{\omega}_{\zeta} = 0$ (with $\zeta = \frac{2}{5}$) for fixed $k_1 = \tfrac{23}{64}$ and after eliminating $k_2$. The vertical and horizontal lines, $\gamma_1$ and $\gamma_2$, are the contours $k_3 = k_1$ and $k_4 = k_1$, respectively. Compare with Fig.~\ref{fig:OmegaNNNEcons} corresponding to the next-nearest neighbor case.}
\label{fig:OmegaExpEcons}
\end{figure}

We analyze the kinematically allowed collisions $\delta(\underline{k})\delta(\underline{\omega}_{\zeta})$ for the dispersion relation in Eq.~\eqref{eq:omega_exp}. A short calculation shows that Eq.~\eqref{eq:omega_exp} can be written as
\begin{equation}
\omega_{\zeta}(k) = -\frac{1}{2} \left(1 + \frac{\sinh(\zeta)}{\cosh(\zeta) - \cos(2\pi k)}\right).
\end{equation}
Again using the momentum conservation $\underline{k} = 0 \mod 1$ and some trigonometric identities, one arrives at the factorization
\begin{equation}
\begin{split}
\underline{\omega}_{\zeta}
&= \frac{1}{2} \sinh(\zeta) \ \underline{\omega}_{\mathrm{bas}} \ \underline{\omega}_{\mathrm{add},\zeta}\\
&\quad \times \left(\prod_{i=1}^4 \left(\cosh(\zeta) - \cos(2\pi k_i)\right)\right)^{-1}
\end{split}
\end{equation}
with the same factor $\underline{\omega}_{\mathrm{bas}}$ as in Eq.~\eqref{eq:omega_bas}, and
\begin{equation}
\label{eq:omega_add_exp}
\begin{split}
\underline{\omega}_{\mathrm{add},\zeta}
&= -\cos(\pi s_{12})^3 + \cos(\pi s_{12})\\
&\quad\quad \times \left(1 + \cosh(\zeta)^2 + \cos(2\pi \Delta k_{12}) \cos(2\pi \Delta k_{34})\right)\\
&\quad - \cosh(\zeta) \left( \cos(2\pi \Delta k_{12}) + \cos(2\pi \Delta k_{34}) \right).
\end{split}
\end{equation}
$\underline{\omega}_{\mathrm{add},\zeta} = 0$ is a cubic equation for $\cos(\pi s_{12})$, which can be solved analytically in closed form or numerically by a few Newton iteration steps. There is only a single real-valued solution, which we (again) denote by $\gamma_{\mathrm{diag}}$ (the context will resolve any ambiguity to the next-neighbor case). Fig.~\ref{fig:OmegaExpEcons} visualizes the contours $\underline{\omega}_{\zeta} = 0$, which resemble the next-nearest neighbor case except that $\gamma_{\mathrm{ellip}}$ is missing and $\gamma_{\mathrm{diag}}$ is slightly distorted. One notices that $\gamma_{\mathrm{diag}}$ and $\gamma_1$ seem to intersect at $(k_3, k_4) = (k_1, 0)$. This is no coincidence, since in the limit $\zeta \to 0$, the equation $\underline{\omega}_{\mathrm{add},\zeta} = 0$ admits a solution $\Delta k_{12} = \Delta k_{34} = \frac12 s_{12}$, which is equivalent to $k_1 = k_3$ and $k_2 = k_4 = 0$. Similarly, $\gamma_{\mathrm{diag}}$ and $\gamma_2$ intersect at $k_1 = k_4$, $k_2 = k_3 = 0$ when $\zeta \to 0$.

The nearest neighbor case~\cite{BoltzmannFermi2012} corresponds to the limit $\zeta \to \infty$: namely, dividing Eq.~\eqref{eq:omega_add_exp} by $\cosh(\zeta)^2$ (which leaves the solutions of $\underline{\omega}_{\mathrm{add},\zeta} = 0$ invariant) and letting $\zeta \to \infty$, only the term $\cos(\pi s_{12})$ remains.


\subsection{Stationary solutions}
\label{sec:StationarySolutions}

In the spatially homogeneous case, the conventional wisdom is that the stationary solutions of the kinetic equation coincide with thermal equilibrium. This should hold also if in~\eqref{eq:Hamiltonian} the lattice $\mathbb{Z}$ is replaced by the $d$-dimensional lattice $\mathbb{Z}^d$. As proved in~\cite{BoltzmannFermi2012}, for a general dispersion relation and in arbitrary dimension the problem of classifying all stationary solutions can be reduced to finding the set of all collision invariants, i.e., solutions to
\begin{equation}
\Phi(k_1) + \Phi(k_2) = \Phi(k_3) + \Phi(k_4)
\end{equation}
on the manifold $\{ (k_1,k_2,k_3,k_4) \,\vert\, \underline{k} = 0 \mod 1,\, \underline{\omega} = 0\}$. The obvious solution reads 
\begin{equation}
\label{eq:CollInvariant}
\Phi(k) = \beta (\omega(k) - \mu_{\sigma}),
\end{equation}
which corresponds to thermal equilibrium. Thus the issue reduces to whether there are further collision invariants. For dimension $d \ge 2$ a proof under fairly general conditions is available~\cite{CollisionalInvariants2006}. For $d = 1$, there could be too few collision channels to reach thermal equilibrium. An example is the nearest neighbor Hubbard chain. There is then no $\gamma_{\mathrm{ellip}}$ and $\gamma_{\mathrm{diag}}$ is linear. As a consequence additional collision invariants can be found. Our numerical simulations indicate that a slight curvature of $\gamma_{\mathrm{diag}}$ suffices to limit the set of collision invariants to the ones listed in~\eqref{eq:CollInvariant}.


\subsection{Integrable models}

The Hubbard chain is integrable for pure $m$-th neighbor hopping models with dispersion relation
\begin{equation}
\label{eq:OmegaMDef}
\omega_m(k) = - \cos(2\pi m k).
\end{equation}
Similar to nearest neighbor hopping ($m = 1$), the energy conservation factorizes as
\begin{equation}
\begin{split}
\underline{\omega}_m &= 4 \, \cos(\pi m (k_3 + k_4)) \\
&\quad \times \sin(\pi m (k_1 - k_3)) \, \sin(\pi m (k_1 - k_4)).
\end{split}
\end{equation}
Accordingly, the collision contours are re-scaled by the factor $m$.

%

There is an infinite number of energy-like conservation laws: Let $g: \mathbb{T} \to \mathbb{R}$ with $g\big(k + \frac{1}{m}\big) = g(k)$ for all $k \in \mathbb{T}$, as well as $g(k) = -g\big(\frac{1}{2 m} - k\big)$. Then
\begin{equation}
\label{eq:GeneralEnergyConservation}
\frac{\ud}{\ud t} \int_{\mathbb{T}} \ud k\, g(k)\, \mathrm{tr}[W(k)] = 0,
\end{equation}
which follows by an appropriate interchange of the integration variables $k_1,\dots,k_4$. Note that $g(k)$ is completely determined by prescribing $g(k)$ for $k \in \big[-\frac{1}{4 m}, \frac{1}{4 m}\big]$.

\section{Numerical Procedure}
\label{sec:Numerics}

\subsection{Contour integrals of the dissipative collision operator}

The following discussion applies to both the next-nearest neighbor and exponential model. Ideally, the numerical discretization of the integration contours (Fig.~\ref{fig:OmegaNNNEcons} and~\ref{fig:OmegaExpEcons}) should preserve the spin and energy conservation laws. These conservation laws result from the interchangeability $k_1 \leftrightarrow k_2$, $k_3 \leftrightarrow k_4$ and the pairs $\{k_1, k_2\} \leftrightarrow \{k_3,k_4\}$. For the contours $\gamma_1$ and $\gamma_2$, we can proceed as in~\cite{BoltzmannFermi2012} using a uniform grid for the $k$ variables. However, the contours $\gamma_{\mathrm{diag}}$ and $\gamma_{\mathrm{ellip}}$ require more sophistication: to adopt the symmetries in the numerical discretization, we first rewrite the dissipative collision evaluated at $k$:
\begin{equation}
\label{eq:rewrite_contour_int}
\begin{split}
&\pi \int_{\mathbb{T}^4} \ud k_1 \ud k_2 \ud k_3 \ud k_4 \, \delta(\underline{k}) \, \delta(\underline{\omega}) \, \delta(k_1-k) \left(\mathcal{A}[W] + \mathcal{A}[W]^*\right)\\
&= \pi \int_{\mathbb{T}^2} \ud \Delta k_{12} \, \ud \Delta k_{34} \int_{\mathbb{T}^2} \ud s_{12} \, \ud s_{34} \,  \delta(s_{12}-s_{34}) \, \delta(\underline{\omega})\\
&\quad\qquad \times \delta\left(s_{12}/2 + \Delta k_{12} - k\right) \left(\mathcal{A}[W] + \mathcal{A}[W]^*\right)\\
&= \pi \int_{\mathbb{T}^2} \ud \Delta k_{12} \, \ud \Delta k_{34} \int_{\mathbb{T}} \ud s_{12} \, \delta(\underline{\omega})\\
&\quad\qquad \times \delta\left(s_{12}/2 + \Delta k_{12} - k\right) \left(\mathcal{A}[W] + \mathcal{A}[W]^*\right),
\end{split}
\end{equation}
where we have used the substitution
\begin{align}
s_{12} &= k_1 + k_2, \quad \Delta k_{12} = \frac{1}{2}(k_1 - k_2),\\
s_{34} &= k_3 + k_4, \quad \Delta k_{34} = \frac{1}{2}(k_3 - k_4).
\end{align}
In the following we are only concerned with the integration along the contour $\gamma_{\mathrm{diag}}$ or $\gamma_{\mathrm{ellip}}$. The $s_{12}$ integral can be eliminated the via $\delta(\underline{\omega})$, namely, 
\begin{equation}
\int_{\mathbb{T}} \ud s_{12} \, \delta(\underline{\omega}) = \abs{\partial s_{12}\,\underline{\omega}}^{-1}.
\end{equation}
Thus the last integral in Eq.~\eqref{eq:rewrite_contour_int} becomes
\begin{equation}
\label{eq:rewrite_contour_int2}
\begin{split}
&\pi \int_{\mathbb{T}^2} \ud\Delta k_{12} \, \ud\Delta k_{34} \abs{\partial s_{12}\,\underline{\omega}}^{-1}\\
&\quad \times \delta\left(s_{12}/2 + \Delta k_{12} - k\right) \left(\mathcal{A}[W] + \mathcal{A}[W]^*\right),
\end{split}
\end{equation}
where $s_{12}$ depends on $\Delta k_{12}$ and $\Delta k_{34}$ via Eq.~\eqref{eq:s12} or $\underline{\omega}_{\mathrm{add},\zeta} = 0$ in Eq.~\eqref{eq:omega_add_exp}, respectively. Numerically, we discretize the integral in~\eqref{eq:rewrite_contour_int2} by a uniform grid:
\begin{equation}
\label{eq:delta_k_discr}
\Delta k_{12} = \frac{j}{n}, \quad j = -\frac{n}{2}, -\frac{n}{2}+1, \dots, \frac{n}{2}-1
\end{equation}
(same for $\Delta k_{34}$) with fixed $n = 128$ in our case. Note that $k_1 \leftrightarrow k_2$ corresponds to $\Delta k_{12} \leftrightarrow -\Delta k_{12}$ and likewise for $\Delta k_{34}$, and that $\{k_1, k_2\} \leftrightarrow \{k_3,k_4\}$ corresponds to $\Delta k_{12} \leftrightarrow \Delta k_{34}$.

So far we have not taken the $\delta$-function in Eq.~\eqref{eq:rewrite_contour_int2} into account, for which we use the following approach: we want to determine the cumulative contribution to the collision operator at the uniform $k$-grid points $k = \frac{j}{n}$, $j = 0, 1, \dots, n-1$. We do not resolve the $\delta$-function exactly; instead, for each term
\begin{equation}
\label{eq:Aterm}
A = \pi \, \abs{\partial s_{12}\,\underline{\omega}}^{-1} \left(\mathcal{A}[W] + \mathcal{A}[W]^*\right)
\end{equation}
evaluated at discretized $\Delta k_{12}$, $\Delta k_{34}$, we first choose $k = \frac{j}{n}$ such that
\begin{equation}
k \le s_{12}/2 + \Delta k_{12} \le k + \frac{1}{n}.
\end{equation}
Then we add $\nu \frac{1}{n} A$ to $\mathcal{C}_\mathrm{d}[W](k)$ and $(1-\nu) \frac{1}{n} A$ to $\mathcal{C}_\mathrm{d}[W](k+\frac{1}{n})$, with $\nu \in \mathbb{R}$ chosen such that
\begin{equation}
\omega(s_{12}/2 + \Delta k_{12}) = \nu\,\omega(k) + (1-\nu)\,\omega\!\left(k+\frac{1}{n}\right).
\end{equation}
By this approach, the numerical scheme preserves the spin and energy conservation laws.

In summary, our numerical method approximates $\mathcal{C}_\mathrm{d}[W](k)$ (and thus $W(k,t)$ for the next time step) at the uniform $k$-grid points $k = \frac{j}{n}$. However, the discretization~\eqref{eq:delta_k_discr} of the terms in Eq.~\eqref{eq:Aterm} requires evaluation of $W(k)$ at $\frac12 s_{12} \pm \Delta k_{12}$ and $\frac12 s_{12} \pm \Delta k_{34}$, which are (in general) no uniform grid points $\frac{j}{n}$. We solve this issue by polynomial interpolation of order $3$ (precomputing divided differences based on $W(k)$ at $k = \frac{j}{n}$).

\subsection{Mollifying the collision operators}
\label{sec:MollificationFormula}

We use the same mollification scheme as in~\cite{BoltzmannFermi2012} to avoid the infinities resulting from $\abs{\partial s_{12}\,\underline{\omega}}^{-1}$ in Eq.~\eqref{eq:Aterm} and the principal value of $1/\underline{\omega}$ in the effective hamiltonian~\eqref{eq:Heff}. Concretely, we replace
\begin{equation}
\label{eq:moll_deriv_omega}
\abs{\partial s_{12}\,\underline{\omega}}^{-1} \to \left(\abs{\partial s_{12}\,\underline{\omega}}^2 + \epsilon^2\right)^{-1/2},
\end{equation}
and for the conservative collision operator
\begin{equation}
\label{eq:moll_inv_omega}
\mathcal{P}\!\left(\frac{1}{\underline{\omega}}\right) \to \frac{\underline{\omega}}{\underline{\omega}^2 + \epsilon^2}
\end{equation}
with finite $\epsilon > 0$. In our case, we use $\epsilon = \frac{1}{50}$ for the simulations in section~\ref{sec:Results}. Note that Eq.~\eqref{eq:moll_inv_omega} becomes an exact identity when taking the limit $\epsilon \to 0$.



\begin{figure}[!ht]
\centering
\includegraphics[width=0.8\columnwidth]{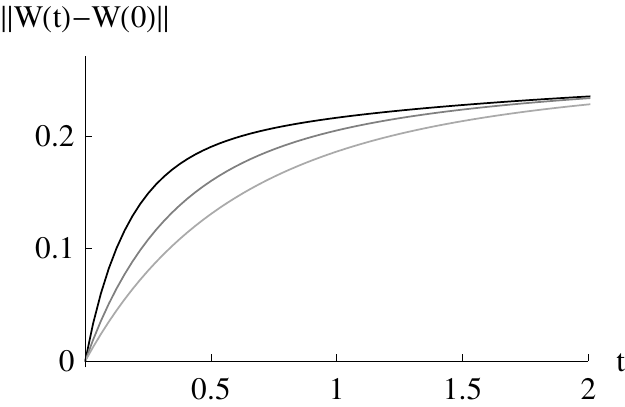}
\caption{Effect of different mollification parameters $\epsilon_1 = \frac{1}{50}$ (upper dark gray curve, used for the simulations in section~\ref{sec:Results}), $\epsilon_2 = \frac{1}{10}$ (middle curve) and $\epsilon_3 = \frac{1}{2}$ (lower light gray curve). The difference to the initial $W(k,0)$ is quantified by the Hilbert-Schmidt norm.}
\label{fig:dW0_epsilon}
\end{figure}

While the mollification parameter $\epsilon$ is required to avoid infinities, it has to be chosen somewhat arbitrarily. We briefly quantify the effect of different values $\epsilon_1 = \frac{1}{50}$, $\epsilon_2 = \frac{1}{10}$ and $\epsilon_3 = \frac{1}{2}$ in Fig.~\ref{fig:dW0_epsilon}. The curves show the Hilbert-Schmidt difference $\lVert W(t) - W(0)\rVert$ between the current and an initial Wigner state (see section~\ref{sec:Results}) up to $t = 2$ (next-nearest neighbor model with $\eta = \frac{1}{2}$). The effects of different mollifications are quite noticeable for the time interval shown. On the other hand, the curves approach each other for larger $t$ since all Wigner states eventually converge to the same thermal equilibrium state. Thus it is reasonable that the asymptotic convergence to equilibrium hardly depends on the mollification.

\subsection{Solving the Boltzmann equation}

Departing from~\cite{BoltzmannFermi2012}, we avoid the Strang splitting technique for treating $\mathcal{C}_{\mathrm{d}}$ and $\mathcal{C}_{\mathrm{c}}$ separately, but simply use the explicit midpoint rule for $\mathcal{C} \equiv \mathcal{C}_{\mathrm{d}} + \mathcal{C}_{\mathrm{c}}$. As advantage, this approach exactly preserves the spin and energy conservation laws. The more laborious time evolution step for $\mathcal{C}_{\mathrm{c}}$ in~\cite{BoltzmannFermi2012} did not show any noticeable differences.

\subsection{Implementation details}

We have implemented the numerical scheme described so far in plain C code, with a custom \texttt{struct} for complex Hermitian $2 \times 2$ matrices (with $4$ \texttt{double} values for the real diagonal entries and the complex $1,2$ entry). The implementation is designed such that the intermediate steps always deal with Hermitian matrices. For example, the commutator $i [A,B]$ and anticommutator $\{A, B\}$ for Hermitian $A, B$ is again Hermitian and can directly be calculated from the matrix entries of $A$ and $B$, without resorting to the products $A B$ or $B A$. Similarly, a custom function calculates the sum of triple products $A B C + C B A$ directly from the matrix entries, which is again Hermitian when $A, B, C$ are.

We use the \emph{MathLink} interface to make the numerical procedures conveniently accessible from Mathematica.

The C implementation comes with a noticeable performance increase: on the same hardware as in~\cite{BoltzmannFermi2012} (Intel Core i7-740QM Processor, 6M cache, 1.73~GHz), a simulation run with the same parameters as in~\cite{BoltzmannFermi2012} now only takes several seconds, as compared to $6\,\mathrm{h}$ for the Mathematica implementation in~\cite{BoltzmannFermi2012}.

\section{Simulation}
\label{sec:Results}

\subsection{Initial Wigner state}
\label{sec:InitialWigner}

\begin{figure}[!ht]
\centering
\includegraphics[width=0.8\columnwidth]{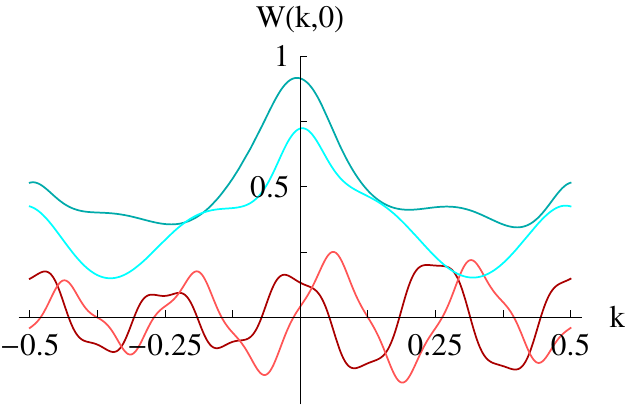}
\caption{(Color online) The initial state $W(k,0)$ used for all simulations in this section. The cyan (upper) curves show the real diagonal entries, and the darker and lighter red curves the real and imaginary parts of the off-diagonal $\lvert\uparrow\rangle \langle\downarrow\rvert$ entry, respectively.}
\label{fig:W0}
\end{figure}

Our goal is to investigate the effects of the different dispersion relations $\omega(k)$ in Fig.~\ref{fig:Omega}. We start with a (rather arbitrary) initial condition $W(k,0)$ shown in Fig.~\ref{fig:W0}. The bright and dark cyan lines represents the real diagonals, and the dark and light red oscillatory functions the real and imaginary part of the off-diagonal $\lvert\uparrow\rangle \langle\downarrow\rvert$ entry, respectively. The eigenvalues of $W(k,0)$ are in the interval $[0,1]$ for each $k \in \mathbb{T}$, as required by the Fermi property. $W(k,0)$ is continuous on $\mathbb{T}$. The analytic formula of $W(k,0)$ can be found in appendix~\ref{sec:AppendixWInitial}.

\begin{figure}[!ht]
\centering
\includegraphics[width=0.9\columnwidth]{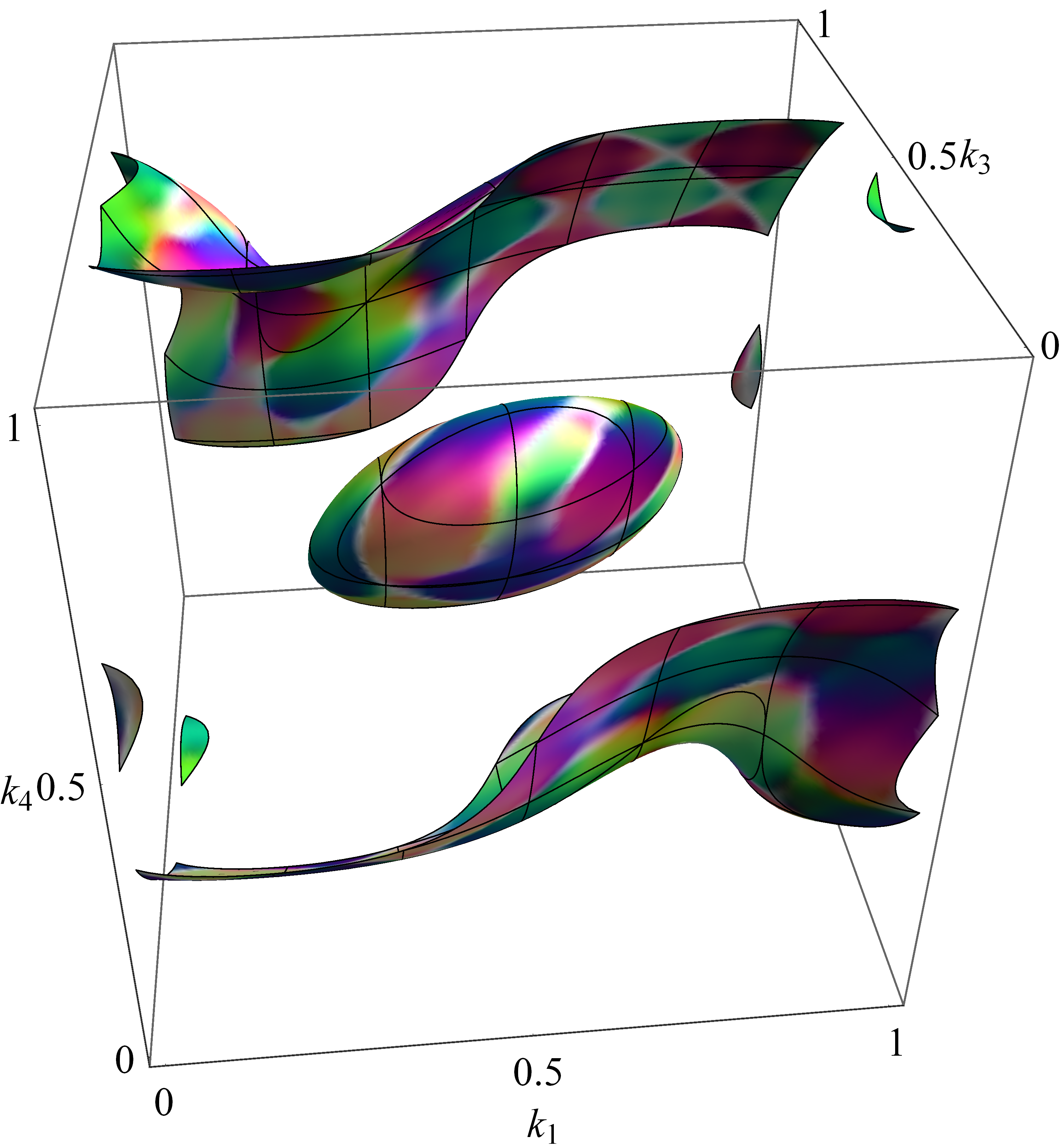}
\caption{(Color online) 3D shape of the $\gamma_{\mathrm{diag}}$ and $\gamma_{\mathrm{ellip}}$ collision manifolds for the next-nearest neighbor model with $\eta = \frac12$ (as in Fig.~\ref{fig:OmegaNNNEcons}). Color encodes the Bloch vector of $\mathcal{A}[W] + \mathcal{A}[W]^*$ for the state $W(k,0)$ shown in Fig.~\ref{fig:W0}.}
\label{fig:collision3D_nnn}
\end{figure}

As illustration, Fig.~\ref{fig:collision3D_nnn} visualizes the 3-dimensional shape of the collision manifolds $\gamma_{\mathrm{diag}}$ and $\gamma_{\mathrm{ellip}}$ for the next-nearest neighbor model with $\eta = \frac{1}{2}$. (Note that Fig.~\ref{fig:OmegaNNNEcons} is the intersection of Fig.~\ref{fig:collision3D_nnn} with the hyperplane $k_1 = \frac{23}{64}$.) To illuminate the effect of the dissipative collision operator $\mathcal{C}_\mathrm{d}$, the colors in Fig.~\ref{fig:collision3D_nnn} encode the Bloch vector of $\mathcal{A}[W] + \mathcal{A}[W]^*$ for the initial state $W(k,0)$, where the red, green and blue colors correspond to the $x$, $y$ and $z$ components of the Bloch vector, respectively.

\subsection{Stationary states}


%
\begin{figure*}[!ht]
\centering
\subfloat[non-thermal stationary state]{
\includegraphics[width=0.8\columnwidth]{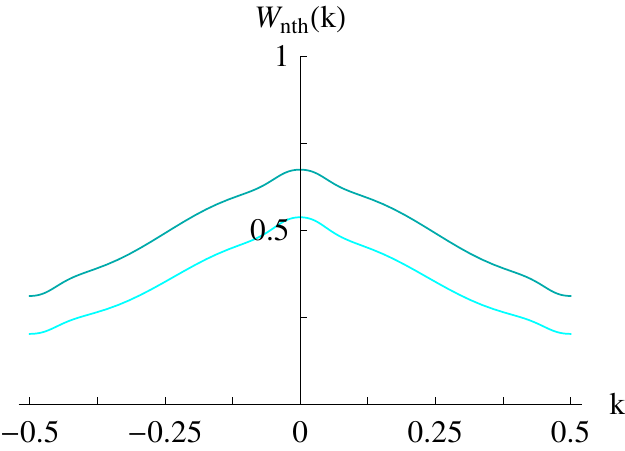}\label{fig:WstationaryNearest}} $\qquad$
\subfloat[thermal equilibrium state for exponential hopping]{
\includegraphics[width=0.8\columnwidth]{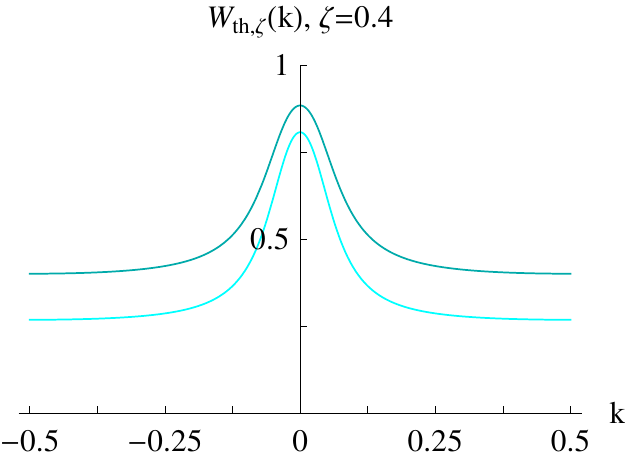}\label{fig:WstationaryExp}} \\
\subfloat[thermal equilibrium state for next-nearest neighbor hopping]{
\includegraphics[width=0.8\columnwidth]{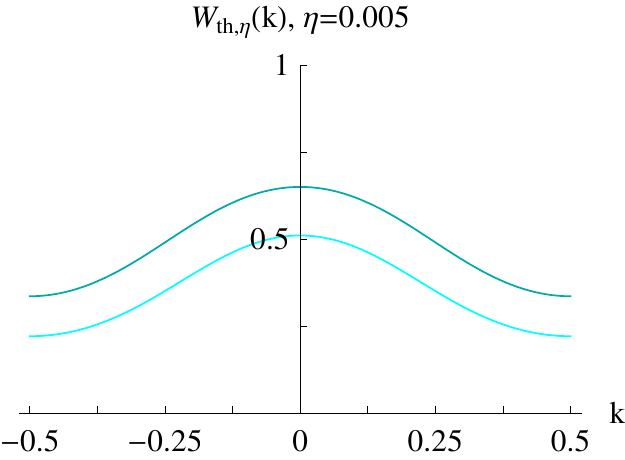}\label{fig:WstationaryNNN1}} $\qquad$
\subfloat[thermal equilibrium state for next-nearest neighbor hopping]{
\includegraphics[width=0.8\columnwidth]{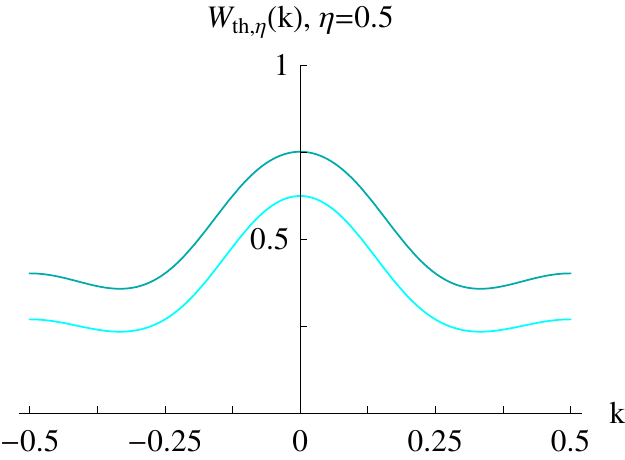}\label{fig:WstationaryNNN2}}
\caption{(Color online) Diagonal matrix entries of the stationary states corresponding to the initial $W(k,0)$ in Fig.~\ref{fig:W0}, for the nearest neighbor hopping model (a), for the exponential hopping model with $\zeta = \frac{2}{5}$ (b), and for the next-nearest neighbor model with $\eta = \frac{1}{200}$ (c) and $\eta = \frac{1}{2}$ (d). The off-diagonal matrix entries are zero.}
\end{figure*}



For the given initial $W(k,0)$, one can obtain the corresponding stationary state (which is different for different dispersion relations) from the conservation laws Eq.~\eqref{eq:SpinConservation}, \eqref{eq:EnergyConservation} and \eqref{eq:TraceConservation}. We will discuss four different models according to Fig.~\ref{fig:Omega}: the nearest neighbor case ($\eta = 0$), the next-nearest neighbor model with a small perturbation $\eta_1 = \frac{1}{200}$ and a larger $\eta_2 = \frac{1}{2}$ (such that the $\gamma_{\mathrm{ellip}}$ collision path opens up), as well as the exponential hopping model with $\zeta = \frac{2}{5}$. The corresponding stationary states are distinct. 



For the nearest neighbor model, the stationary solution is a non-thermal state of the form
\begin{equation}
\label{eq:stationary_nonthermal}
W_{\mathrm{nth}}(k) = \sum_{\sigma \in \{\uparrow,\downarrow\}} \left( \mathrm{e}^{f(k) - a_\sigma} + 1 \right)^{-1} \lvert\sigma\rangle\langle\sigma\rvert,
\end{equation}
i.e., a real diagonal $k$-dependent matrix, where the function $f$ satisfies the symmetry property $f(k) = -f(\frac{1}{2} - k)$. $f$ is obtained numerically, and the corresponding $W_{\mathrm{nth}}(k)$ shown in Fig.~\ref{fig:WstationaryNearest}. The cyan lines visualize the diagonal entries.

With a perturbation $\eta \neq 0$ in the next-nearest neighbor model, the stationary states become thermal states of the form
\begin{equation}
\label{eq:thermal_eta}
W_{\mathrm{th},\eta}(k) = \sum_{\sigma \in \{\uparrow, \downarrow \}} \left(\mathrm{e}^{\beta (\omega_\eta(k) - \mu_\sigma)} + 1\right)^{-1} \lvert\sigma \rangle \langle \sigma\rvert.
\end{equation}
Figs.~\ref{fig:WstationaryNNN1} and~\ref{fig:WstationaryNNN2} visualize $W_{\mathrm{th},\eta}(k)$ calculated from the initial $W(k,0)$. Compared to $f(k)$, the term $\beta\,\omega_\eta(k)$ lacks the symmetry $f(k) + f(\frac{1}{2}-k) = 0$. Note that even for $\eta \to 0$, in general $W_{\mathrm{th},\eta}(k)$ does not converge to $W_{\mathrm{nth}}$.

The numerically obtained values of $\beta$ and $\mu_\sigma$ in Eq.~\eqref{eq:thermal_eta} are summarized in the following table:
\begin{center}
\begin{tabular}{l|ccc}
& $\beta$ & $\mu_\uparrow$ & $\mu_\downarrow$\\
\hline
$\eta_1 = 0.005$ & $0.650$ & $0.949$ & $0.061$ \\
$\eta_2 = 0.5$ & $0.752$ & $0.972$ & $0.176$ \\
\end{tabular}
\end{center}



The stationary state of the exponential hopping model is a thermal equilibrium state of the form
\begin{equation}
\label{eq:thermal_exp}
W_{\mathrm{th},\zeta}(k) = \sum_{\sigma \in \{\uparrow, \downarrow \}} \left(\mathrm{e}^{\beta (\omega_{\zeta}(k) - \mu_\sigma)} + 1\right)^{-1} \lvert\sigma \rangle \langle \sigma\rvert
\end{equation}
as shown in Fig.~\ref{fig:WstationaryExp}. The corresponding parameters are $\beta = 1.00$, $\mu_\uparrow = -1.00$ and $\mu_\downarrow = -1.60$. The peak around $k = 0$ becomes sharper when $\zeta$ decreases.
%

\subsection{Exponential convergence, fast and slow motion}

\begin{figure*}[!ht]
\centering
\subfloat[entropy increase]{
\includegraphics[width=0.8\columnwidth]{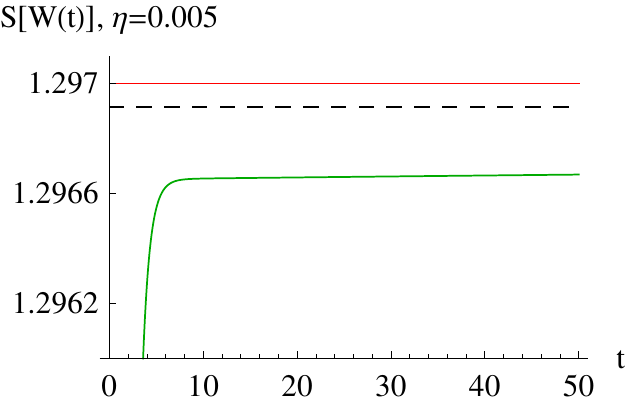}
\label{fig:entropy_conv_eta1}
} $\qquad$
\subfloat[convergence of the off-diagonal entries]{
\includegraphics[width=0.8\columnwidth]{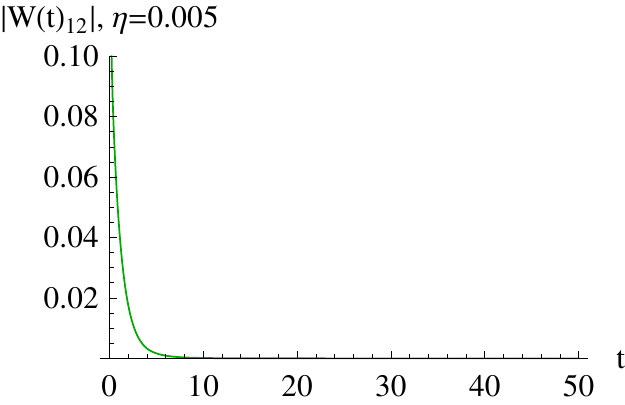}
\label{fig:offdiag_conv_eta1}}
\caption{(Color online) Entropy increase for the next-nearest neighbor model with small $\eta = \frac{1}{200}$ (green curve). The red curve shows the entropy of the corresponding equilibrium state, and the dashed black curve the entropy of the stationary nearest neighbor state. The entropy grows very slowly after $t \simeq 10$.}
\end{figure*}

We pick the entropy as representative measure of convergence to stationarity. In our numerical simulations, we observe exponential convergence, i.e., the entropy difference
\begin{equation}
S[W_\mathrm{st}] - S[W(t)] \simeq \mathrm{e}^{-\kappa t}
\end{equation}
for large times $t$, where $W_\mathrm{st}$ denotes the respective stationary state. The following table summarizes the decay rates $\kappa$ obtained from a least-squares fit in logarithmic representation:
\begin{center}
\begin{tabular}{r|c|c|c|c}
& nearest & $\eta_1 = 0.005$ & $\eta_2 = 0.5$ & $\zeta = 0.4$\\
\hline
$\kappa$ & 0.852 & 0.001 & 0.0676 & 0.0530
\end{tabular}
\end{center}
One notices that the convergence rate is highest for the nearest neighbor model, and lowest for the next-nearest neighbor model with small $\eta_1 = \frac{1}{200}$. We investigate the latter case in more detail. The green line in Fig.~\ref{fig:entropy_conv_eta1} shows a closeup of the entropy $S[W(t)]$ in dependence of $t$, and the red line the entropy value $S[W_{\mathrm{th,\eta}}(k)] = 1.297$ of the corresponding stationary state. For comparison, the dashed line is the entropy of the non-thermal stationary state ($\eta = 0$). One notices that the entropy grows much faster when $t \le 10$ and then reaches a plateau, where it approaches the asymptotic red line very slowly. This observation suggests the following dynamical picture: In the phase space for~\eqref{eq:BoltzmannEquation} there is the slow manifold consisting of Wigner functions of the form~\eqref{eq:stationary_nonthermal}. A general initial state, $W$, will rapidly move towards the slow manifold, and will arrive there at a state $W(t_*)$, where in general $W(t_*) \neq W_{\mathrm{nth}}$. From there on there is an effective dynamics on the slow manifold with initial Wigner function $W(t_*)$.
This can be seen in Fig.~\ref{fig:offdiag_conv_eta1}. To obtain the evolution equation in the slow manifold, we treat the off-diagonal entries as small perturbation,
\begin{equation}
W(k) = W^\mathrm{D}(k) + \delta \, W^\mathrm{OD}(k)
\end{equation}
with $0 < \delta \ll 1$, $W^\mathrm{D}(k)$ the diagonal part and $W^\mathrm{OD}(k)$ the off-diagonal part. The effective dynamics for the state is driven by
\begin{equation}
\mathcal{C}[W^\mathrm{D} + \delta \, W^\mathrm{OD}](k).
\end{equation}
Since the conservative collision operator $\mathcal{C}_\mathrm{c}$ in Eq.~\eqref{eq:Cc} is defined by a commutator, it holds that
\begin{equation}
\mathcal{C}_\mathrm{c}[W](k,t) = \mathcal{O}(\delta).
\end{equation}
Thus the Boltzmann differential equation is to zero-th order in $\delta$ governed by the dissipative part coupling the $\uparrow\uparrow$ and $\downarrow\downarrow$ correlation functions:
\begin{equation}
\frac{d}{\ud t} W^\mathrm{D}(k,t) = \mathcal{C}_\mathrm{d}^\mathrm{D}[W^\mathrm{D}](k,t) + \mathcal{O}(\delta),
\end{equation}
where
\begin{equation}
\label{eq:Cd_diag}
\begin{split}
& \mathcal{C}_\mathrm{d}^\mathrm{D}[W_{\uparrow \uparrow}](k,t) = \pi \int_{\mathbb{T}^3} \ud k_2 \ud k_3 \ud k_4 \delta(\underline{k}) \delta(\underline{\omega})\\
& \times \left( \tilde{W}_{1,\uparrow \uparrow} \tilde{W}_{2,\downarrow \downarrow} W_{3,\uparrow \uparrow} W_{4,\downarrow \downarrow} - W_{1,\uparrow \uparrow} W_{2,\downarrow \downarrow} \tilde{W}_{3,\uparrow \uparrow} \tilde{W}_{4,\downarrow \downarrow} \right).
\end{split}
\end{equation}
The differential equation for $W_{\downarrow \downarrow}$ is given by interchanging $\uparrow \uparrow$ and $\downarrow \downarrow$ in Eq.~\eqref{eq:Cd_diag}. We suspect that this is the effective equation for the slow-motion dynamics.

\begin{figure*}[!ht]
\centering
\subfloat[exponential decay rate in dependence of $\eta$]{
\includegraphics[width=0.8\columnwidth]{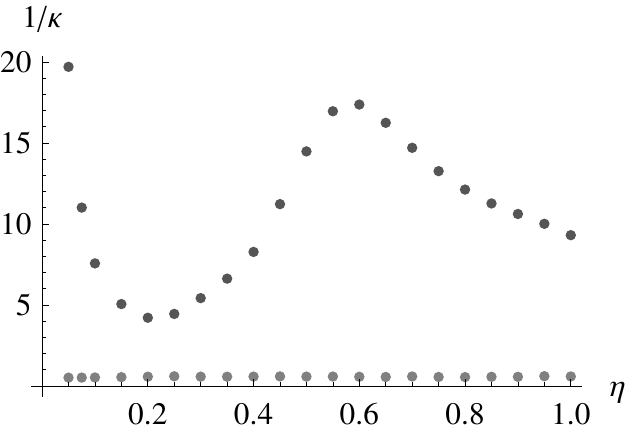}} $\qquad$
\subfloat[entropy convergence for $\eta = \frac{1}{2}$]{
\includegraphics[width=0.8\columnwidth]{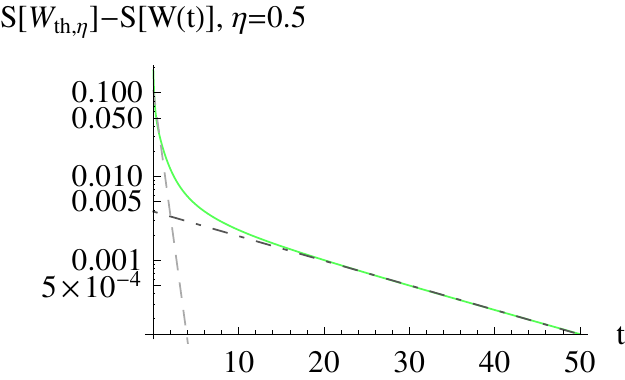}}
\caption{(Color online) (a) Inverse exponential decay rate $1/\kappa$ of the entropy difference in dependence of $\eta$ (next-nearest neighbor model), obtained from a least squares fit as exemplified in (b). The upper dark gray points in (a) correspond to the asymptotic decay rate for large $t$ (dark dot-dashed line in (b)), and the lower light gray points to the initial decay rate at $t = 0$ (light dashed line in (b)).}
\label{fig:entropy_conv}
\end{figure*}

The concept of different dynamical regimes is supported by Fig.~\ref{fig:entropy_conv}: The dark gray points represent the inverse asymptotic decay rates $1/\kappa$ for the next-nearest neighbor model in dependence of $\eta$. For comparison, the light gray points show the initial decay rates at $W(k,0)$. One observes that initial and asymptotic decay rates are clearly separated.


\section{Conclusions}
\label{sec:Conclusions}
On the level of the Boltzmann-Hubbard equation one can easily destroy integrability by going beyond the next-nearest neighbor hopping. The structure of the kinetic equation is not touched, but through modifying $\omega$ one changes the set of allowed collisions. The consequences on the dynamics are in accordance with text book wisdom. In the integrable case the collision rule has a high symmetry and, while there is still exponential convergence and non-zero entropy production, in general one reaches a nonthermal state of the form~\eqref{eq:stationary_nonthermal}. Any tiny modification of $\omega$ restores the physically expected thermalization to the Fermi-Dirac diagonal Wigner function. For large modifications we find again exponential fast convergence. However, for a small perturbation of $\omega$, we clearly demonstrated two time scales, a rapid convergence to quasi-stationarity and a slow convergence to thermal equilibrium.

Our model is fairly simple, but serves as an example where the approach (and non-approach) to thermal equilibrium can be studied in detail.
\appendix
\section{Analytic formula of $W(k,0)$} \label{sec:AppendixWInitial}
For the sake of reproduceability, the analytic formula of the initial Wigner state $W(k,0)$ used in the simulations (section~\ref{sec:InitialWigner}) reads as follows:
\begin{multline}
W_{\uparrow\uparrow}(k,0) = \left(\mathrm{e}^{\frac{1}{2} \left(\cos(2 \pi k) - \cosh(2/5)\right)^{-1} \sinh(2/5) + \frac{1}{2}}+1\right)^{-1} \\
 + \frac{1}{432} \bigg(18 \cos\big(\pi(6 k + 1/7)\big) - 14 \cos\big(6 \pi(k-1/7)\big) \\
 + 27 \left(\cosh (1)-\cos (4 \pi k)\right)^{-1} \Big(\mathrm{e}^{-3/5}\cos(2 \pi k) \\
 + \cos(4 \pi k) - \mathrm{e}^{2/5} \cos (6 \pi k) + \mathrm{e}^{-1}\Big)
\bigg),
\end{multline}
\begin{multline}
W_{\uparrow\downarrow}(k,0) = \frac{1}{54} \Big(9 \sin\!\left(\mathrm{e}^{8 i \pi k}\right) - (1+i) \cos\big(6 \pi(k-1/7)\big)\\
+ 6\,(1-i) \sin\big(\pi(3 k + 1/14)\big) \sin\big(3 \pi(k - 1/7)\big)\Big)
\end{multline}
together with
\begin{equation}
W_{\downarrow\uparrow}(k,0) = W_{\uparrow\downarrow}(k,0)^{*},
\end{equation}
and
\begin{multline}
W_{\downarrow\downarrow}(k,0) \\
= \left(\mathrm{e}^{\frac{1}{2} \left(\cos(2 \pi k) - \cosh(2/5)\right)^{-1} \sinh(2/5) + \frac{11}{10}}+1\right)^{-1} \qquad \\
 + \frac{1}{432} \bigg(14\cos\big(6 \pi(k - 1/7)\big) - 18 \cos\big(\pi(6\,k + 1/7)\big)\\
 + 27 \left(\cosh(3/2) - \cos(4\pi k)\right)^{-1} \Big(\mathrm{e}^{-11/10}\cos(2\pi k) \\
 + \cos(4\pi k) - \mathrm{e}^{2/5} \cos(6\pi k) - \mathrm{e}^{-3/2}\Big)\bigg).
\end{multline}


\begin{thebibliography}{10}

\bibitem{Essler2010}
F.~H.L. Essler, H.~Frahm, F.~G{\"o}hmann, A.~Kl{\"u}mper, and V.~E. Korepin.
\newblock {\em {The one-dimensional Hubbard model}}.
\newblock Cambridge University Press, 2010.

\bibitem{HubbardModelPhysics1995}
D.~Baeriswyl, D.~K. Campbell, J.~M.P. Carmelo, F.~Guinea, and E.~Louis,
  editors.
\newblock {\em {The Hubbard model: its physics and its mathematical physics}}.
\newblock Nato Science Series B. Springer, 1995.

\bibitem{DiffusionBallisticTransport2009}
J.~Sirker, R.~G. Pereira, and I.~Affleck.
\newblock {Diffusion and Ballistic Transport in One-Dimensional Quantum
  Systems}.
\newblock {\em Phys. Rev. Lett.}, 103:216602, 2009.

\bibitem{ConservationIntegrabilityTransport2011}
J.~Sirker, R.~G. Pereira, and I.~Affleck.
\newblock {Conservation laws, integrability, and transport in one-dimensional
  quantum systems}.
\newblock {\em Phys. Rev. B}, 83:035115, 2011.

\bibitem{QuantumLiquids2012}
A.~Imambekov, T.~Schmidt, and L.~Glazman.
\newblock {One-dimensional quantum liquids: Beyond the Luttinger liquid
  paradigm}.
\newblock {\em Rev. Mod. Phys.}, 84:1253--1306, 2012.

\bibitem{BoltzmannFermi2012}
Martin L.~R. F\"urst, Christian~B. Mendl, and Herbert Spohn.
\newblock {Matrix-valued Boltzmann equation for the Hubbard chain}.
\newblock {\em Phys. Rev. E}, 86:031122, 2012.

\bibitem{TransportWeaklyInteracting2012}
Ch.~Bartsch and J.~Gemmer.
\newblock {Boltzmann-type approach to transport in weakly interacting
  one-dimensional fermionic systems}.
\newblock {\em Phys. Rev. E}, 85:041103, 2012.

\bibitem{ErdosSalmhoferYau2004}
L.~Erd\H{o}s, M.~Salmhofer, and H.-T. Yau.
\newblock {On the quantum Boltzmann equation}.
\newblock {\em J. Stat. Phys.}, 116:367--380, 2004.

\bibitem{NotToNormalOrder2009}
J.~Lukkarinen and H.~Spohn.
\newblock {Not to normal order -- Notes on the kinetic limit for weakly
  interacting quantum fluids}.
\newblock {\em J. Stat. Phys.}, 134:1133--1172, 2009.

\bibitem{MeiLukkarinenSpohnPrep}
P.~Mei, J.~Lukkarinen, and H.~Spohn.
\newblock {The Hubbard-Boltzmann equation}.
\newblock {\em in preparation}, 2012.

\bibitem{CollisionalInvariants2006}
H.~Spohn.
\newblock {Collisional invariants for the phonon Boltzmann equation}.
\newblock {\em J. Stat. Phys.}, 124:1131--1135, 2006.

\end{thebibliography}
\end{document}